\documentclass[12pt]{article}
\begin{document}

\def \d {{\rm d}}
\def \A {{\rm A}}
\def \B {{\rm B}}
\def \CU {{\cal U}}
\def \CV {{\cal V}}

\title{Expanding impulsive gravitational waves}

\author{J. Podolsk\'y\thanks{E--mail: {\tt Podolsky@mbox.troja.mff.cuni.cz}}
\\
\\ Department of Theoretical Physics, Charles University,\\
V Hole\v{s}ovi\v{c}k\'ach 2, 18000 Prague 8, Czech Republic.\\
\\
and J. B. Griffiths\thanks{E--mail: {\tt J.B.Griffiths@Lboro.ac.uk}} \\ \\
Department of Mathematical Sciences, Loughborough University \\
Loughborough, Leics. LE11 3TU, U.K. \\ } 

\date{\today}

\maketitle
\begin{abstract}
We explicitly demonstrate that the known solutions for expanding impulsive
spherical gravitational waves that have been obtained by a ``cut and paste''
method may be considered to be impulsive limits of the Robinson--Trautman
vacuum type~N solutions. We extend these results to all the generically
distinct subclasses of these solutions in Minkowski, de~Sitter and
anti-de~Sitter backgrounds. For these we express the solutions in terms of a
continuous metric. Finally, we also extend the class of spherical shock
gravitational waves to include a non-zero cosmological constant.
\\ \\ Pacs {04.20.Jb, 04.30.Nk}
\end{abstract}

\section{Introduction}
In a classic article in 1972, Penrose \cite{Pen72} presented a novel method
for constructing impulsive gravitational waves in a Minkowski background by
cutting the space-time along a null hypersurface and then re-attaching the
two pieces with a suitable warp. He showed that two types of solution could
be constructed by this technique --- the familiar impulsive
{\sl pp}-waves, and spherical waves that are described as impulsive limits
of the Robinson--Trautman waves with spherical wavefronts.

The latter class (and related classes) of spherical impulsive waves have
recently received considerable attention. The explicit solution written in a
continuous coordinate system has been given by Nutku and Penrose
\cite{NutPen92} and Hogan \cite{Hogan93}, \cite{Hogan94}. This solution has
been interpreted \cite{NutPen92} as representing either the snapping of a
cosmic string (identified by a deficit angle) in the exterior region
generating a spherical impulsive gravitational wave, or as having an
expanding string generated behind the gravitational wave. The collision (and
breaking) of a pair of cosmic strings can also be described in this way.

A solution with a similar interpretation has also been described by
Bi\v{c}\'ak and Schmidt~\cite{BicSch89}. In this case, however, two null
particles recede from a common point generating an impulsive spherical
gravitational wave. In this solution, which was obtained as a limiting case of
a solution with a boost-rotation symmetry, there is an expanding cosmic
string along the axis of symmetry separating the two particles. It was later
pointed out \cite{Bicak90} that this situation is equivalent to that of the
splitting of an infinite cosmic string as described by Gleiser and
Pullin~\cite{GlePul89} or, rather, of two semi-infinite cosmic strings
approaching at the speed of light and separating again at the instant at which
they collide.

A similar solution in a space-time with a non-vanishing cosmological constant
has been given by Hogan \cite{Hogan92}. This is a natural extension of the
solution \cite{Hogan93} to describe a spherical impulsive wave in a de~Sitter
background, although the solution is also valid when the cosmological constant
is negative.

However, despite the statement in Penrose's initial article \cite{Pen72},
the above solutions have never been explicitly related to the
Robinson--Trautman solutions. Using the original \cite{RobTra60} or
Garc\'{\i}a--Pleba\'nski \cite{GarPle81} coordinate systems, it is found that
the impulsive limit of the Robinson--Trautman solutions introduces a term in
the metric tensor which is quadratic in the Dirac $\delta$-function. The
difficulty in treating singular expressions like the square of the
$\delta$-distribution can in principle be solved by considering a careful
regularization (at least for some special class of regularizing functions)
and then performing an appropriate distributional limit of the corresponding
well behaved approximation. Such an approach has recently been used, for
example, in the context of the equation of geodesic deviation for impulsive
{\sl pp}-waves \cite{Stein98}. Moreover, the recently developed Colombeau
theory on nonlinear generalized functions \cite{Col84}, \cite{Col92} provides
a new and rigorous mathematical framework for a possible consistent treatment
of such nonlinear operations with distributions. Here, we have not used the
full rigour of the Colombeau theory. Nevertheless, we have demonstrated how
the problematic singular terms can be removed using an explicit (necessarily
discontinuous) coordinate transformation, thus producing a continuous form of
the metric.

The first purpose of this paper is to use this transformation to explicitly
demonstrate that the impulsive gravitational waves generated by the ``cut and
paste'' method may be considered to be impulsive Robinson--Trautman
solutions. This is achieved in section~3.

The second purpose of this paper is to complete the full class of such
solutions by taking impulsive limits of all the nine generically distinct
subclasses \cite{BicPod98} of the Robinson--Trautman type~N vacuum
solutions with a cosmological constant. These describe impulsive spherical
gravitational waves in de~Sitter or anti-de~Sitter backgrounds, and are
presented in section~4. The complete family of these solutions is expressed
in a coordinate system in which the metric is explicitly continuous. This
unifies and generalises previously known results. We will also show that the
limits of the distinct classes with different $\epsilon$ (representing all
permitted Gaussian curvatures of spacelike surfaces) are locally equivalent.
This degeneracy is identical to that observed for the impulsive limits of the
Kundt class of solutions as demonstrated recently~\cite{Podol98b}.

Finally, we will also consider the natural extension of the above methods to
construct solutions which represent expanding spherical gravitational shock
(step) waves. A shock wave of Robinson--Trautman type~N has been constructed
by Nutku \cite{Nutku91} for the case in which the cosmological constant
vanishes. In section~5, we will generalise this to the case of non-zero
$\Lambda$ in which the shock wave propagates into a de~Sitter or
anti-de~Sitter background.

\section{The type N Robinson--Trautman solutions}

Before considering impulsive limits, it is first appropriate to review some
general aspects of the type N Robinson--Trautman family of vacuum solutions
that will be needed below. These solutions are well known  (see \cite{KSMH80}).
In a familiar coordinate system they can be written in the form
 $$ \d s^2 =2{r^2\over P^2}\,\d\zeta\,\d\bar\zeta +2\d u\,\d r
-2\left[\epsilon+r(\log P)_u-{\Lambda\over6}r^2\right]\d u^2, $$
 where $2\epsilon(u)$ can be interpreted as the Gaussian curvature of the
2-surfaces $2P^{-2}\d\zeta\d\bar\zeta$, and $\Lambda$ is the cosmological
constant. In these cases, it is always possible to use the coordinate
transformation $u=g(\tilde u)$, $r=\tilde r/\dot g$, where $\dot g=\d
g/\d\tilde u$, to put $\epsilon=-1,0,+1$.
 It then remains to solve the Robinson--Trautman equation $\Delta\log
P=2\epsilon$, where $\Delta=2P^2\partial_\zeta\partial_{\bar\zeta}$. A general
solution of this equation can be expressed as
 \begin{equation}
 P=\big(1+\epsilon F\bar F\big)
\big(F_\zeta\bar F_{\bar\zeta}\big)^{-1/2},
 \label{P}
 \end{equation}
 where $F=F(u,\zeta)$ is an arbitrary complex function of $u$ and $\zeta$,
holomorphic in $\zeta$.
The only non-vanishing component of the Weyl tensor is given by
$$\bar\Psi_4 =-{P^2\over2r}F_\zeta
\left[{1\over F_\zeta}\big(\log
F_\zeta\big)_{u\zeta}\right]_\zeta\ .$$

It is natural to introduce the invariant classification of the type~N
Robinson--Trautman vacuum solutions given in \cite{GarPle81}. However, we will
adopt the slightly modified notation introduced in \cite{BicPod98}. This
classification is based on two parameters: $\Lambda$ and $\epsilon$, each of
which may be either zero, positive or negative. Together, these indicate a
total of nine distinct subclasses denoted as $RTN(\Lambda,\epsilon)$.

An alternative representation of this $RTN(\Lambda,\epsilon)$ class of
solutions which is more suitable for interpretation has been given by
Garc\'{\i}a D\'{\i}az and Pleba\'nski \cite{GarPle81}. Their metric can be
expressed in terms of an arbitrary complex function $f(\xi,u)$ which is
holomorphic in the complex spatial coordinate $\xi$ by
 \begin{eqnarray}
 \d s^2 &=& 2v^2\d\xi\d\bar\xi +2\psi\,\d u\,\d v  \nonumber\\
&&\
+2v\bar \A\,\d\xi\,\d u +2v\A\,\d\bar\xi\,\d u +2(\A\bar\A+\psi\B)\d u^2,
 \label{GP}
 \end{eqnarray}
 where
 \begin{eqnarray}
 \A&=&\epsilon\xi-vf, \nonumber\\
 \B&=&-\epsilon +{v\over2}(f_\xi+\bar f_{\bar\xi})+{\Lambda\over6}v^2\psi,
\nonumber\\
 \psi&=&1+\epsilon\xi\bar\xi. \nonumber
 \end{eqnarray}
 The complete family of solutions described by this metric can now be denoted
as $RTN(\Lambda,\epsilon)[f]$ in which the free parameters $\Lambda$ and
$\epsilon$ and the arbitrary function $f(\xi,u)$ are identified explicitly.

The transformation which relates the above two forms of the
Robinson-Trautman solutions is (see \cite{BicPod98})
 \begin{eqnarray}
 \xi&=& F(\zeta,u) =\int f(\xi(\zeta,u),u)\d u\ , \nonumber\\
 v&=& {r\over1+\epsilon F\bar F}\ . \nonumber
 \end{eqnarray}
 (If $f$ is independent of $\xi$, it is necessary to put $\xi=F=\zeta+\int
f(u)\d u$). The Weyl tensor component (using a rescaled tetrad) is now given
simply by
 \begin{equation}
 \bar\Psi_4 =-{1\over2r}\,f_{\xi\xi\xi}.
 \label{Psi4}
 \end{equation}
From this, it is clear that the metric (\ref{GP}) represents only a conformally
flat Minkowski or (anti-) de~Sitter background when $f$ is no more than
quadratic in $\xi$. In all other cases it describes a radiative vacuum
space-time of type~N~\cite{BicPod98b}.

\section{Impulsive waves in a Minkowski background}

Although the Robinson--Trautman solutions naturally describe ``spherical''
gravitational waves, it had previously been observed \cite{Pen72},
\cite{Hogan93} that there are difficulties in considering the impulsive limit
in this family since the metric in the standard coordinate system is
quadratic in the $\delta$ function, and this gives rise to mathematical
problems.

Let us first concentrate on the simplest subclass of type~N
Robinson--Trautman solutions of the form (\ref{GP}) in which
$\Lambda=\epsilon=0$ and consider the impulsive limit as $f(\xi,u)\to
f(\xi)\delta(u)$, i.e. $RTN({\Lambda=0},{\epsilon=0})[{f=f\delta}]$. In this
case the line element approaches the form
 \begin{eqnarray}
 \d s^2 &=& 2v^2\d\xi\d\bar\xi +2\d u\,\d v  \nonumber\\
&&\
-2v^2\bar f\,\delta(u)\,\d\xi\,\d u
-2v^2f\,\delta(u)\,\d\bar\xi\,\d u \nonumber\\
&&\  +\big[2v^2f\bar f\,\delta^2(u)
+v(f_\xi+\bar f_{\bar\xi})\delta(u)\big] \d u^2.
 \label{RTN00}
 \end{eqnarray}
In this form of the metric, the above difficulty is seen explicitly.
The question arises whether a form of the metric exists in which
these problematic terms do not appear. In considering a
transformation to such a form, it may be observed that, to remove
the $\delta$-function components in (\ref{RTN00}), a discontinuous
transformation is necessary. In fact, all the singular components can
be removed by making the transformation
 \begin{eqnarray}
\xi&=& Z+\Theta(U)\left[ -Z+h(Z)+{U\over2V}{\alpha\bar\beta\over{\cal A}}
\right], \nonumber\\
\bar\xi&=& \bar Z+\Theta(U)\left[ -\bar Z+\bar h(\bar Z)
+{U\over2V}{\bar\alpha\beta\over{\cal A}} \right], \nonumber\\
v&=& V+\Theta(U)\left[ -V+V{{\cal A}\over|\alpha|}\right], \nonumber\\
u&=& U+\Theta(U)\left[ -U+U{|\alpha|\over{\cal A}}\right],
 \label{trans0}
 \end{eqnarray}
 where $\Theta(U)$ is the Heaviside step function, $h(Z)$ is an arbitrary
function related to $f$ by $f(\xi)\equiv h(Z)-Z$ evaluated on $U=0$,
$\alpha=h'$, $\beta=h''/h'$, ${\cal A}=1-U\beta\bar\beta/4V$, and the
derivative with respect to $Z$ is denoted by a prime. With this, the line
element (\ref{RTN00}) takes the form
 \begin{equation}
\d s^2= 2V^2\left|\d Z+{U\Theta(U)\over2V}\bar H\,\d\bar Z \right|^2
+2\d U\,\d V,
 \label{L=0e=0}
 \end{equation}
where $H=\beta'-{1\over2}\beta^2=h'''/h'-{3\over2}(h''/h')^2$. In this form,
the metric is explicitly continuous with the impulsive component $\delta(U)$
appearing only in the curvature tensor. In fact, this is the known metric for an
expanding impulsive spherical gravitational wave in a Minkowski background
obtained by Nutku and Penrose \cite{NutPen92} and Hogan \cite{Hogan93} using
the Penrose \cite{Pen72} ``cut and paste'' method\footnote
       {In this paper we denote the impulsive wave surface by {$U=0$},
interchanging the coordinates $U$ and $V$ in \cite{NutPen92}, \cite{Hogan93}
and elsewhere.}.

The transformation (\ref{trans0}) is an explicit combination of the
transformations used in the ``cut and paste'' method in \cite{Hogan93}. It is
remarkable that, in this form, it exactly removes all the $\delta$-function
components in (\ref{L=0e=0}). Conversely, the transformation (\ref{trans0})
applied inversely to the Nutku--Penrose--Hogan metric (\ref{L=0e=0}) yields
the same problematic metric (\ref{RTN00}) as is obtained as the impulsive
limit of the $RTN({\Lambda=0},{\epsilon=0})$ solution. In this sense, the
line element (\ref{L=0e=0}) may be considered to be equivalent to the
impulsive limit of the type~N Robinson--Trautman solution (\ref{GP}) for the
case in which $\Lambda=\epsilon=0$.

It may be commented that the discontinuity in the transformation
(\ref{trans0}) is necessary to remove the impulsive components in the metric
(\ref{RTN00}). This is very similar to the analogous transformation
between the Brinkmann and the continuous forms of the impulsive {\sl pp}-wave
metric that has recently been investigated elsewhere \cite{KunSte99}, in
which the distributional and continuous forms of the metric have rigorously
been shown to be equivalent. However, in the present case, the rigorous
calculations are considerable more complicated. This is currently under
investigation.

We now move on to describe other spherical impulsive waves in a Minkowski
background. These may be considered as belonging to the family
$RTN({\Lambda=0},\epsilon=\pm1)[{f=f\delta}]$.

In this case, it is first convenient to transform the line element (\ref{GP})
with $f=f(\xi)\delta(u)$ by putting
 $$ w=\psi v. $$
 With this, (\ref{GP}) takes the form
 \begin{eqnarray}
 \d s^2 &=& 2{w^2\over\psi^2}\d\xi\d\bar\xi +2\d u\,\d w -2\epsilon\d u^2
\nonumber\\ &&\
-2{w^2\over\psi^2}\bar f\,\delta(u)\,\d\xi\,\d u
-2{w^2\over\psi^2}f\,\delta(u)\,\d\bar\xi\,\d u \nonumber\\
&&\  +\Big[2{w^2\over\psi^2}f\bar f\,\delta(u)
-2\epsilon{w\over\psi}f\bar\xi -2\epsilon{w\over\psi}\bar f\xi 
 +w(f_\xi+\bar f_{\bar\xi})
\Big] \delta(u) \d u^2.
 \label{RTN0e}
 \end{eqnarray}
 This clearly includes the above case (\ref{RTN00}) when $\epsilon=0$, but it
also includes the two additional cases when $\epsilon=\pm1$.

The line element (\ref{RTN0e}) again contains the square of the
$\delta$-function, but this problem can similarly be removed by making the
more complicated coordinate transformation
 \begin{eqnarray}
\xi&=& Z+\Theta(U)\left[ -Z+\xi_p  \right], \nonumber\\
\bar\xi&=& \bar Z+\Theta(U)\left[ -\bar Z+\bar\xi_p \right], \nonumber\\
w&=& V+\Theta(U)\left[ -V+w_p \right], \nonumber\\
u&=& U+\Theta(U)\left[ -U+u_p \right],
 \label{transe}
 \end{eqnarray}
where (assuming $\epsilon\not=0$ only)
 \begin{eqnarray}
\xi_p&=& -\frac{\epsilon}{2\bar\eta}\left[ (\CU-\epsilon\CV)
 \mp\sqrt{(\CU-\epsilon\CV)^2+4\epsilon\eta\bar\eta} \right], \nonumber\\
w_p&=& -(\CU-\epsilon\CV)-\frac{4\epsilon\eta\bar\eta}
{(\CU-\epsilon\CV)\mp\sqrt{(\CU-\epsilon\CV)^2+4\epsilon\eta\bar\eta}}\ ,
 \nonumber\\
u_p&=& -\epsilon\CU-\frac{2\eta\bar\eta}
{(\CU-\epsilon\CV)\mp\sqrt{(\CU-\epsilon\CV)^2+4\epsilon\eta\bar\eta}}\ ,
 \label{transe2}
 \end{eqnarray}
with
 \begin{eqnarray}
\CU&=& AV-DU,   \nonumber\\
\CV&=& BV-EU,   \nonumber\\
\eta&=& CV-FU,  \nonumber\\
 \label{transe3}
 \end{eqnarray}
and
 \begin{eqnarray}
&&A= \frac{\sqrt2}{p|h'|},\qquad
B= \frac{|h|^2}{\sqrt2\, p|h'|},\qquad
C= \frac{h}{ p|h'|},   \nonumber\\
&&D= \frac{\sqrt2}{|h'|}\left\{
\frac{p}{4} \left|\frac{h''}{h'}\right|^2+\epsilon
\left[1+\frac{Z}{2}\frac{h''}{h'}+\frac{\bar Z}{2}\frac{\bar h''}{\bar h'}
\right]\right\},\nonumber\\
&&E= \frac{|h|^2}{\sqrt2|h'|}\bigg\{
\frac{p}{4}\left|\frac{h''}{h'}-2\frac{h'}{h}\right|^2  
+\epsilon\left[ 1+\frac{Z}{2}
\left(\frac{h''}{h'}-2\frac{h'}{h}\right)+\frac{\bar Z}{2}
\left(\frac{\bar h''}{\bar h'}-2\frac{\bar h'}{\bar h}\right)
\right]\bigg\}\  ,\nonumber\\
&&F= \frac{h}{|h'|}\bigg\{
\frac{p}{4}\left(\frac{h''}{h'}-2\frac{h'}{h}\right)
\frac{\bar h''}{\bar h'}  
+\epsilon\left[1+
 \frac{Z}{2}\left(\frac{h''}{h'}-2\frac{h'}{h}\right)
+\frac{\bar Z}{2}\frac{\bar h''}{\bar h'}\right]\bigg\}\ ,
 \label{transe4}
 \end{eqnarray}
 where $f$ is related to $h(Z)$ through the identification $f\equiv\xi_p-Z$
evaluated on $U=0$.

With this transformation the line element (\ref{RTN0e}) finally takes the form
 \begin{equation}
\d s^2= {2V^2\over p^2}
\left| \d Z+{U\Theta(U)\over2V}\,p^2\bar H\,\d\bar Z \right|^2
+2\d U\,\d V -2\epsilon\,\d U^2,
 \label{L=0en0}
 \end{equation}
 where $p=1+\epsilon Z\bar Z$. This is exactly the solution for an expanding
spherical impulsive gravitational wave in a Minkowski background that was
obtained by Hogan \cite{Hogan94}. It is explicitly continuous
and  obviously reduces to (\ref{L=0e=0}) when $\epsilon=0$.

The transformation (\ref{transe})--(\ref{transe4}) is evidently
rather complicated. However, it reduces to an identity when $U<0$
so that the metric  (\ref{RTN0e}) with $f=0$ goes directly to
(\ref{L=0en0}). On the other hand, for $U>0$ the transformation
(\ref{transe})--(\ref{transe2}) is inverse to
 \begin{eqnarray}
\CU&=& \frac{w}{\psi}-\epsilon u,   \nonumber\\
\CV&=& \frac{\xi\bar\xi w}{\psi}-u, \nonumber\\
\eta&=& \frac{\xi w}{\psi},
 \label{inv}
 \end{eqnarray}
and brings the same ``background'' $RTN({\Lambda=0},\epsilon)[f=0]$
metric given by (\ref{RTN0e}) to an explicit Minkowski form
$$ \d s^2 = 2\d\eta\d\bar\eta  - 2\d \CU\,\d \CV\ . $$
This metric is related to (\ref{L=0en0}) with $U>0$ by
(\ref{transe3})--(\ref{transe4}). Note that the transformation
(\ref{transe})--(\ref{transe4}) cannot simply be applied when
$\epsilon=0$. The reason is that (\ref{transe2}) is not a
correct inverse relation to (\ref{inv}) in this case. However,
using the correct one  $\xi=\eta/\CU$, $v=\CU$,
$u=\eta\bar\eta/\CU-\CV$ with (\ref{transe3})--(\ref{transe4}),
we get exactly the transformation (\ref{trans0}) after a trivial
rescaling $\xi\to\sqrt2\,\xi$, $v\to v/\sqrt2$ and $u\to\sqrt2\,u$.

Let us here also present some useful relations valid for the
above transformations. First, it can easily be shown that
$ \sqrt{(\CU-\epsilon\CV)^2+4\epsilon\eta\bar\eta}=|w|$, so that
we have to consider only the upper signs in Eqs. (\ref{transe2}) when
$w>0$ and the lower signs when $w<0$. This naturally explains an
apparent ambiguity of the transformation.

Secondly, it immediately follows from Eqs. (\ref{inv}) that
$\eta\bar\eta-\CU\CV\equiv u(w-\epsilon u)$. Substituting
from (\ref{transe3}) and using the identities
 \begin{eqnarray}
C\bar C-AB &=& 0,   \nonumber\\
F\bar F-DE  &=& -\epsilon,   \nonumber\\
AE+BD-C\bar F-\bar CF&=&1,  \nonumber
 \end{eqnarray}
we get an interesting relation
 \begin{equation}
 u(w-\epsilon u)= U(V-\epsilon U),
 \label{ident}
 \end{equation}
which will be important in the next section.

Note finally that in the Robinson--Trautman metrics with {\it general}
profile function $f(\xi,u)$, the 2-surfaces spanned by $\zeta$ have
constant curvature which may be zero, positive or negative
according to the value of $\epsilon$.  However, it can
be shown that, for the limiting case of {\it impulsive} waves,
these three cases are equivalent. Indeed, the simple coordinate transformation
 \begin{eqnarray}
\xi &\to& \xi\,{v\over v-\epsilon u}\ , \nonumber\\
v &\to& v-\epsilon u\ , \nonumber\\
u &\to& u\,{\psi v-\epsilon u\over v-\epsilon u}\ ,
 \label{equi}
 \end{eqnarray}
 takes the solution (\ref{RTN00}) $RTN({\Lambda=0},\epsilon=0)[f=f\delta]$ to
(\ref{RTN0e}) $RTN({\Lambda=0},\epsilon)[f=f\delta]$. This degeneracy of the
impulsive limit of the Robinson--Trautman solutions is similar to that of the
Kundt class of solutions for non-expanding impulsive waves as shown in
\cite{Podol98b}.

\section{Impulsive waves in a de~Sitter or anti-de~Sitter background}

It is fairly straightforward to generalise the above solution (\ref{L=0e=0})
to the case of an expanding spherical gravitational wave in a de~Sitter, or
anti-de~Sitter background. This was achieved by Hogan \cite{Hogan92}
expressing the metric in the similar continuous form:
 \begin{equation}
\d s^2= {1\over(1+{\Lambda\over6}UV)^2} 
 \bigg[  2V^2 \left| \d Z+{U\Theta(U)\over2V}\bar H\,\d\bar Z
\right|^2 + 2\d U\,\d V \bigg].
 \label{Ln0e=0}
 \end{equation}
 By applying the transformation (\ref{trans0}), we can relate (\ref{Ln0e=0})
exactly to the metric (\ref{RTN00}) multiplied by a conformal factor
$(1+{\Lambda\over6}uv)^{-2}$ since (\ref{trans0}) has the nice property that
$uv=UV$. Performing the further transformation
 \begin{equation}
\tilde v={v\over1+{\Lambda\over6}uv}
 \label{konf1}
 \end{equation}
 and dropping the tildes, we immediately obtain the impulsive subclass of the
metric (\ref{GP}) which represents the class of solutions denoted as
$RTN(\Lambda,{\epsilon=0})[f=f\delta]$ in which $\Lambda$ may be zero,
positive or negative.

It is now natural to look for the alternative form of this class of
solutions for which both $\Lambda$ and $\epsilon$ may be non-zero.
This corresponds to a different coordinate
representation of (\ref{Ln0e=0}) similar to the alternative form
(\ref{L=0en0}) of the solution (\ref{L=0e=0}) for the case when $\Lambda=0$. A
general solution satisfying these properties would then be valid for all
possible values of $\Lambda$ and $\epsilon$ thus covering all nine generically
different subclasses of the Robinson--Trautman vacuum type~N solutions.

This general form of the metric in a continuous coordinate system may be
written in the form
 \begin{equation}
\d s^2={1\over[1+{\Lambda\over6}U(V-\epsilon U)]^2} 
\,\bigg[ {2V^2\over p^2}
\left| \d Z+{U\Theta(U)\over2V}\,p^2\bar H\,\d\bar Z \right|^2
+2\d U\,\d V -2\epsilon\,\d U^2 \bigg]. 
 \label{Ln0en0} 
 \end{equation}
 As in the above case, this can be derived using the transformation
(\ref{transe})--(\ref{transe4}) from the metric (\ref{RTN0e}) multiplied by
the conformal factor
 \begin{equation}
 {1\over\big[1+{\Lambda\over6}u(w-\epsilon u)\big]^2}\ ,
 \end{equation}
 considering the identity (\ref{ident}). Further, this can be written
in the form of the most general impulsive metric of the Robinson--Trautman
type
 \begin{eqnarray}
 \d s^2 &=& 2{w^2\over\psi^2}\d\xi\d\bar\xi +2\d u\,\d w
+\Big({\Lambda\over3}w^2-2\epsilon\Big)\d u^2
\nonumber\\ &&\
-2{w^2\over\psi^2}\bar f\,\delta(u)\,\d\xi\,\d u
-2{w^2\over\psi^2}f\,\delta(u)\,\d\bar\xi\,\d u \nonumber\\
&&\  +\Big[2{w^2\over\psi^2}f\bar f\,\delta(u)
-2\epsilon{w\over\psi}f\bar\xi -2\epsilon{w\over\psi}\bar f\xi 
 +w(f_\xi+\bar f_{\bar\xi})
\Big] \delta(u) \d u^2
 \label{RTNLe}
 \end{eqnarray}
 by putting
 \begin{eqnarray}
\tilde w&=& {w\over1+{\Lambda\over6}u(w-\epsilon u)}\ , \nonumber\\
\tilde u&=& \int{\d u\over1-\epsilon{\Lambda\over6}u^2}\ ,
 \label{trans2}
 \end{eqnarray}
 and dropping the tildes. The metric (\ref{RTNLe}) describes all possible
subclasses $RTN(\Lambda,\epsilon)[f=f\delta]$ in which $\Lambda$ and
$\epsilon$ may be zero, positive or negative. It reduces to the forms
(\ref{RTN0e}) when $\Lambda=0$ and further to (\ref{RTN00}) when
$\epsilon=0$.

The above argument demonstrates that the continuous form of the metric given
in (\ref{Ln0en0}) covers all possible subclasses. In terms of continuous
coordinates, the new metric (\ref{Ln0en0}) clearly reduces to the three
previously known forms (\ref{L=0en0}), (\ref{Ln0e=0}) and (\ref{L=0e=0}) when
$\Lambda=0$ or $\epsilon=0$ or both.

However, it should again be stressed that in the impulsive cases for each
$\Lambda$, the forms for different $\epsilon$ are equivalent. This can easily
be shown by rewriting the impulsive subclasses
$RTN(\Lambda,{\epsilon=0})[f=f\delta]$ of the metric (\ref{GP}) using the
transformation (\ref{konf1}) in the form which is conformal to (\ref{RTN00})
with the conformal factor
$(1+{\Lambda\over6}uv)^{-2}$. Now the transformation (\ref{equi}) brings the
metric to the form (\ref{RTN0e}) of $RTN(\Lambda=0,\epsilon)[f=f\delta]$
multiplied by the conformal factor $[1+{\Lambda\over6}u(w-\epsilon u)]^{-2}$.
Finally, using (\ref{trans2}) we obtain (\ref{RTNLe}), i.e.
$RTN(\Lambda,\epsilon)[f=f\delta]$.

\section{Shock waves in de~Sitter and anti-de~Sitter backgrounds}

It can be observed that the above solutions describing an impulsive
gravitational wave can easily be adapted to describe shock (step) waves.
Spherical shock waves belonging to the Robinson--Trautman type~N class have
been explicitly constructed by Nutku \cite{Nutku91}. These cover the case
in which the cosmological constant vanishes so that the shock propagates
into a Minkowski background. This was obtained initially using a
modification of Penrose's method of identification with warp \cite{Pen72} in
which the space-time behind the wavefront is not flat. When the metric
functions are smooth, these solutions are equivalent to the
Robinson--Trautman vacuum solutions.

It may be noted that there are also other solutions for shock waves with a
spherical wave-front which propagate into a Minkowski background
\cite{AleGri96}. These are algebraically general, indicating the presence
of backscattering (wave tails). However, here we will concentrate on the
Robinson--Trautman solutions which take a similar form to the impulsive
solutions described above. In fact we can easily extend the known solution
to the case in which the cosmological constant is non-zero.

We may again start with the metric (\ref{GP}) which describes the
Robinson--Trautman type~N vacuum solutions in the Garc\'{\i}a--Pleba\'nski
coordinate system \cite{GarPle81}. Using the transformation
 \begin{eqnarray}
\xi&=& F(\zeta,u)\ , \nonumber\\
v&=& {\tilde r\over\sqrt{F_\zeta \bar F_{\bar\zeta}}}\ , \nonumber
 \end{eqnarray}
 we obtain the Nutku form of the metric
 \begin{eqnarray}
 \d s^2 &=& 2\tilde r^2\d\zeta\d\bar\zeta +2P\,\d u\,\d\tilde r
 \label{Nutku} \\
&& +2\tilde r P_\zeta\,\d\zeta\,\d u
   +2\tilde r P_{\bar\zeta}\,\d\bar\zeta\,\d u
+\Big({\Lambda\over3}\tilde r^2P^2-2\epsilon\Big)\d u^2, \nonumber
 \end{eqnarray}
 where $P(\zeta,\bar\zeta,u)$ is given by (\ref{P}) in terms of $F(\zeta,u)$
which satisfies
 $$ {\partial F\over\partial u}=f\big(\xi=F(\zeta,u),u\big), $$
 where $f$ is the arbitrary complex function which appears in the metric
(\ref{GP}).

The spherical shock wave arises when the function $F(\zeta,u)$ is of the form
 $$ F=F\big(\zeta+u\Theta(u)\big). $$
 With this, the metric (\ref{Nutku}) generalises the Nutku solution
\cite{Nutku91} to the case of non-vanishing cosmological constant.

In view of (\ref{Psi4}), the simplest spherical shock wave in Minkowski,
de~Sitter and anti-de~Sitter backgrounds is generated by the function $f$
which is cubic in $\xi$. It is natural to consider
 $$ f(\xi,u) =-\Theta(u)\,\xi^3\ , $$
 which corresponds to
 $$ F(\zeta,u) ={1\over\sqrt2\sqrt{\zeta+u\Theta(u)}}\ , $$
i.e.
 $$ P=\sqrt{2|\zeta+u\Theta(u)|}\,(2|\zeta+u\Theta(u)|+\epsilon)\ . $$
 Although the metric (\ref{GP}) is discontinuous in this case, the metric
(\ref{Nutku}) is explicitly continuous and there is a discontinuity in the
Weyl tensor component~(\ref{Psi4}).

\section{Conclusions}

By constructing explicit transformations, we have shown that the known
solutions for spherical impulsive gravitational waves constructed by the
``cut and paste'' method may be considered to be impulsive limits of the
type~N Robinson--Trautman family of solutions.

We have considered these impulsive limits for all the nine generically
different subclasses of the Robinson--Trautman type~N solutions
$RTN(\Lambda,\epsilon)$ and presented a continuous coordinate system unifying
and generalizing previously known forms for arbitrary $\Lambda$ and
$\epsilon$. However, as in the non-expanding type~N Kundt class, the
solutions for different values for $\epsilon$ are found to be equivalent in
these impulsive limits.

Finally, we have presented a family of exact solutions for expanding spherical
shock waves of Robinson--Trautman type in conformally flat backgrounds.
This generalises a previously known solution in a Minkowski background
to include the equivalent solution in de~Sitter or anti-de~Sitter
backgrounds.

\section*{Acknowledgments}

This work was supported by a visiting fellowship from the Royal Society and,
in part, by the grant GACR-202/99/0261 of the Czech Republic.

\end{document}